\input harvmac.tex
\def\g{\gamma}

\def\be{{\beta}}
\def\s{\sigma}

\def\avd#1{\overline{#1}}

\def\H{{\cal H}}

\lref\gib{M. Serva and G. Paladin, Phys. Rev. Lett. {\bf 70 }, 105 (1993)}
\lref\lib{A. Crisanti, G. Paladin and A. Vulpiani, 
{\it Products of random matrices in statistical physics} 
 (Series in Solid State Sciences 104,  Springer-Verlag) (1993)}
\lref\ose{V.I. Oseledec,  Trans. Moscow. Math. Soc. {\bf 19}, 197 (1968)}
\lref\mat{T.D. Schultz, D.C. Mattis and E.H. Lieb, Rev. Mod. 
Phys. {\bf 36}, 856 (1964)}
\lref\sha{R. Shankar and G. Murthy, Phys. Rev. B {\bf 36}, 536 (1987)}
\lref\ser{M. Serva, Exact solution of a 2d random Ising model (submitted)}
\lref\serv{M. Serva, 2$d$ Ising model with layers of quenched spins 
(submitted) }
\lref\ons{L. Onsager, Phys. Rev. {\bf 65}, 117 (1944)}
\lref\vill{J. Villain, J. Phys. C {\bf 10}, L537 (1977)}
\lref\parI{G. Parisi, Phys. Lett. {\bf 73A}, 203 (1979)}
\lref\parII{G. Parisi, J. Phys. A {\bf 13}, L115 (1980)}
\lref\parIII{G. Parisi, J. Phys. A {\bf 13}, 1101 (1980)}
\lref\mezI{M. M\'ezard, G. Parisi, N. Sourlas, G. Toulouse and M. A. Virasoro,
J. Phys. {\bf 45}, 843 (1984)}
\lref\mezII{M. M\'ezard and M. A. Virasoro, J. Phys. {\bf 46}, 1293 (1985)}
\lref\nosI{G. Paladin, M. Pasquini and M. Serva, J. Phys. France I {\bf 4},
1597 (1994)}
\lref\nosII{M. Pasquini, G. Paladin and M. Serva, Phys. Rev. E {\bf 51}, 
2006 (1995)}
\lref\nosIII{G. Paladin, M. Pasquini and M. Serva, Int. J. Mod. Phys. B 
{\bf 9}, 399 (1995)}
\lref\nosIV{G. Paladin, M. Pasquini and M. Serva, J. Phys. France I {\bf 5},
337 (1995)}
\lref\nosV{S. Scarlatti, M. Serva and M. Pasquini, J. Stat. Phys. {\bf 80},
337 (1995)}
\lref\tou{G. Toulouse, Commun. Phys. {\bf 2}, 115 (1977)}
\lref\zoo{S. Kobe and T. Klotz, Phys. Rev. E {\bf 52}, 5660 (1995)}
\lref\che{H.-F. Cheung and W. McMillan, J. Phys. C {\bf 16}, 7027 (1983)}
\lref\fis{M. E. Fisher and W. Selke, Phys. Rev. Lett. {\bf 44}, 1502 (1980)}
\lref\kou{F. Koukiou, Europhys. Lett. {\bf 7}, 297 (1992)}
\lref\sau{L. Saul and M. Kardar, Phys. Rev. E {\bf 48}, R3221 (1993)}
\lref\tho{M. F. Thorpe and D. Beeman, Phys. Rev. B, {\bf 14}, 188 (1976)} 
\lref\van{J. L. Van Hemmen and R. G. Palmer, J. Phys. A, {\bf 15}, 3881 
(1982)}

\nfig\entropy{Zero temperature entropy $S_0$ as function of $p$ 
at $\g<1$, $\g=1$ and $\g>1$.
In the last case $S_0$ becomes positive for $p>\tilde{p}$, 
where $\tilde{p}<{3\over4}$, and  $S_0(p={3\over4})\simeq 0.01$.}

\nfig\phases{$p - T$ phase diagram at different $\g$: 
a) $\g=0.8$; b) $\g=1$; c) $\g=1.2$; d) $\g=3$.}

\nfig\cal{Specific heat $C$ as function of $T$ for $\g=1.2$ and $p=0.82$.
The first peak, located at $T\simeq0.299$, is referred 
to an antiferromagnetic transition, while the others, magnified in the boxes 
($T\simeq0.349$ and $T\simeq0.545$, to ferromagnetic transitions. The circles
in the boxes represent our numerical data.}

\nfig\pla{A typical realization of the system. The full lines represent
the $+1$ ferromagnetic bonds, while the dashed lines are the
$-\g$ antiferromagnetic bonds.
The 'a' elementary plaquettes are frustrated, at difference with
the 'b' plaquettes.}

\vskip 1.truecm\noindent
\centerline{\bf 2D FRUSTRATED ISING MODEL WITH FOUR PHASES}
\vskip 1.4truecm\noindent
\centerline{M. Pasquini and M. Serva}
\vskip .5truecm
\centerline{\it Dipartimento di Matematica and Istituto Nazionale
Fisica della Materia, 
Universit\`a dell'Aquila}
\centerline{\it I-67010 Coppito, L'Aquila, Italy}
\vskip 1.6truecm
\centerline{ABSTRACT}
\vskip .4truecm
In this paper we consider a $d=2$ random Ising system 
on a square lattice with nearest neighbour interactions.
The disorder is short range correlated and  
asymmetry between the vertical and the horizontal direction
is admitted.
More precisely,
the vertical bonds are supposed to be non random 
while the horizontal bonds alternate: one row of all non random 
horizontal bonds is followed by one row where they
are independent dichotomic random variables.
We solve the model using an approximate approach that replace the
quenched average with an annealed average under the constraint that
the number of frustrated plaquettes is keep fixed and equals
that of the true system. 
The surprising fact is that for some choices of the parameters
of the model there are three second order phase transitions
separating four different phases: 
antiferromagnetic, glassy-like, ferromagnetic and paramagnetic.

\hfill\break
\noindent
PACS NUMBERS: 05.50.+q, 02.50.+s 
\vfill\eject

\newsec{Introduction}

Mean field spin glasses models have been studied and 
deeply understood both from a static and a dynamic point of view
and key words like replica symmetry breaking, aging and 
ultrametricity,  have become of very wide use in 
statistical mechanics of disordered systems
\refs{\parI\parII\parIII\mezI{-}\mezII}.
The reason why spin glasses have attracted  so much attention 
is probably more a consequence of the many successful applications to 
biological modeling (neural networks, immune system, adaptive evolution)
then to their original scope limited to
the description of disordered materials.
For this reason and may be for objective technical difficulties
most of the typical features which are very well
established for the mean field models
have not been found out for short range spin glasses.
For example, it is commonly believed that a finite temperature 
glassy phase only exists for $d \ge 3$ spin glasses
while in $d=2$ one has only the paramagnetic phase.
This is an almost surely true statement if one consider $d=2$
spin system with independent bonds \refs{\che\fis\kou\sau\van{-}\tho}
and with 
vertical-horizontal symmetry but may be it is a false statement 
if one consider $d=2$ spin asymmetric systems
with correlated disorder.
For example, in models with layered disorder
the existence of a low temperature phase seems to be 
an established fact \refs{\sha\ser{-}\serv}, nevertheless, 
one may think that these models are pathological
since layered disorder is somehow a long range correlated disorder.

In this paper we consider a $d=2$ Ising system where there is 
both a short range correlation of the disorder
and an asymmetry between vertical and horizontal direction.
The specific interaction we chose is not motivated by
a deep physical insight but it is merely dictated by technical reasons. 
Nevertheless, the model is not very artificial and the disorder correlation
is limited to the fact that frustrated plaquettes always 
are present in near couples while the 
asymmetry only lies in a difference of strength
of vertical and horizontal bonds.

We solve the model using an approximate approach that replaces the
quenched average with an annealed average under the constraint that
the number of frustrated plaquettes is keep fixed and equals
that of the true system. 
The surprising fact is that for some choices of the parameters
of the models one can find four different phases.

The paper is organized as follows.

In section 2, after a brief overlook of the constrained annealing,
we introduce our model with a particular attention to
the concept of frustration; then we write
the partition function with constrained frustration and the relative
free energy.

In section 3 we derive the solution of the model, obtaining
an expression for the free energy that can be computed 
via numerical methods.
Moreover, the ground state energy is exactly found.

In section 4 the conditions that yield to second order phase transitions
are derived.

In section 5 we describe the various behaviours of the model,
showing a total of four distinct phases, three almost conventional
(high temperature paramagnetic phase, ferromagnetic
and antiferromagnetic phases at low temperature) and a fourth 
paramagnetic phase that we guess to have a 'glassy' nature.

In section 6 we present our conclusions.

\newsec{Constrained annealing}

The model is defined on a square $d=2$ lattice and 
the interaction is supposed to
be effective only between nearest neighbours. The number of spins is 
$N=LM$ where $M$ is the number of columns of the lattice 
and $L$ is the number of rows.

The vertical bonds are supposed to be non random and one can assume
without loss of generality that they equal $1$ while
the horizontal bonds alternate; one row of all non random 
horizontal equal $1$ bonds is followed by one row where they
are independent dichotomic random variables which 
equal $1$ with probability $p$ and equal the negative value 
$-\g$ with probability $1-p$ (see \pla).

It follows that the Hamiltonian of our model can be written as:
\eqn\H{
H_N=-\sum_{i=1}^L \sum_{j=1}^M \left (\s_{i,j} \s_{i+1,j}
+ J_{i,j}\s_{i,j} \s_{i,j+1} \right) 
} 
where $\s_{i,j}=\pm 1$ is the spin in the site located by the
$i$-th row and the $j$-th column while the $J_{i,j}$ are the horizontal 
bonds which equal $1$ when $i$ is even and are defined by

\eqn\coupling{
J_{i,j} = \left\{ {\eqalign{
    & \quad 1 \ \qquad 
		\quad {\rm with \ probability}\ 1-p\cr 
    & -\g \qquad  
		\quad {\rm with \ probability} \quad p}} \right. 
} 
when $i$ is odd.

The model is parameterized by $\g>0$ and $p$
and, in general, it is random, except in the two limit cases $p \to 0$
and $p \to 1$. 
In the first limit case $p=0$ all the couplings equals the unity 
and, therefore, we have the pure $d=2$ Ising model \ons .
In the second limit case $p=1$ the model is also not random,
but while all the vertical couplings equal the unity,
the horizontal couplings alternate one row in which they are all positive 
and equal the unity to one row in which they are all negative and equal $-\g$.
In this second limit the model can be solved by standard transfer
matrix methods \ons\ and it shows a low temperature magnetic phase;
for $\g<1$ this low temperature phase is ferromagnetic
while for $\g>1$ there is horizontal antiferromagnetic order 
and vertical ferromagnetic order between
the spins. 
For the sake of simplicity
we will call hereafter this complicated magnetic phase simply
the antiferromagnetic phase.
Finally, the special choice $p=1$, $\g=1$ corresponds 
the so-called 'fully frustrated model' \vill which is also not random
and has a transition only at $T=0$

In order to explain the nature of our approximation,
let us first recall that the elementary unit for frustration is the
plaquette. If the product of the signs of the bonds around a 
plaquette is negative the plaquette is frustrated, otherwise, 
the plaquette is unfrustrated.
In our Ising model, only the sign of the random variable 
$J_{i,j}$ with odd $i$ can be negative, therefore 
the two square plaquettes which share this bond 
are frustrated if this bond is effectively negative
(e.g. 'a' plaquettes in \pla)
and they are unfrustrated if it is positive ('b' plaquettes).
As a consequence of this definition
of elementary frustration, we may define the total frustration 
of the system $\phi_N$ 
as the rate of frustrated plaquettes. 
In our model
\eqn\fru{\phi_N=
{2\over\ N} \sum_{i=1}^{L} \sum_{j=1}^M {1-J_{i,j} \over 1+\g}
}
This quantity equals $p$ in average, furthermore the 
strong law of large numbers assures that 
$\phi_N \to p$
with probability $1$ in the thermodynamic limit.

We are far from being able to solve the quenched model,
nevertheless we think that the qualitative behaviour of the system
is captured by the above definition of total frustration 
( \tou , for a more general definition see \zoo ).
Therefore, our proposal is to consider an annealed approximation
where $\phi_N$ is constrained to
coincide, in the thermodynamic limit, 
with the quenched total frustration $p$.
This model corresponds to averaging $Z$ only over the realizations
of the disorder with total frustration $p$.
We not only believe that the approximated model
has the same qualitative features of the quenched one,
but it is also in good quantitative agreement with it.
In fact, our experience is that constrained annealing is a 
really powerful tool for estimating the free energy of
disordered systems \refs{\gib\nosI\nosII\nosIII\nosIV{-}\nosV}.
We would like also to stress that the fixed frustration model
can be also seen as an independent model
where the bonds as well as the spins are allowed to arrange themselves
in order to minimize the free energy provided they satisfy
the global frustration constraint.

In order to obtain the free energy of the fixed frustration model
we follow the general method (\gib, \nosIII).
We must first define the generalized partition function
\eqn\z{
Z_N (\be,\g,\mu)=\avd{ 
\sum_{\sigma} \exp \left[ -\be H_N + 
\mu N(\phi_N -p )  \right]} 
}
where $\be={1\over T}$ is the inverse temperature, and the average 
is over all realizations of the couplings $J_{i,j}$,
than we obtain the free energy of the constrained annealed model as
\eqn\free{
f(\be,\g) = - \min_\mu \lim_{N \to \infty} {1\over{N \be}} \ln Z(\be,\g,\mu)
}
where the $N\to \infty$ limit means that both $M$ and $L$ must
tend to the same limit.
In fact, the minimization over $\mu$ automatically selects 
the realizations of the disorder for which $\phi_N = p$
in the thermodynamic limit.

\newsec{Solution of the model}

The generalized partition function is a sum of a product of 
randomly independent variables, therefore, we can write
$$
Z_N (\be,\g,\mu)=
\sum_{\sigma} \prod_{i=1}^L \prod_{j=1}^M 
\avd{ \exp 
\left[ \be \s_{i,j} \s_{i+1,j} +
\be J_{i,j} \s_{i,j} \s_{i,j+1}
+\mu (2{1-J_{i,j} \over 1+\g }-p)
\right] }
$$
The average
can be now easily performed, obtaining:
\eqn\zb{
Z_N (\be,\g,\mu)=\exp \left[ {N\over4} \Big(
 \ln\left(4p(1-p)\right) -2 (2p-1) \mu +a \Big) \right]
\sum_{\{\s\}} \exp\left[-\be \tilde{H}_N  \right]
}
where we have introduced the effective hamiltonian 
$$
\tilde{H}_N=-\sum_{i=1}^L \sum_{j=1}^M \left( \s_{i,j} \s_{i+1,j}
+\tilde J_i \s_{i,j} \s_{i,j+1} \right)
$$
The new effective horizontal bonds $ \tilde{J}_i$
are not random, they are all equal in the same row,
and they alternate two possible values in different rows;
in fact, one has $ \tilde{J}_i = 1$  when $i$ is even
and  $ \tilde{J}_i = {b\over 2\be}$ when $i$ is odd.
The constants $a$ and $b$ are 
\eqn\ab{
\left\{ {\eqalign{
    & a= \ln \left[ 
\cosh \left(\be {1+\g\over2} - \mu+ 
{1\over2} \ln {{1-p}\over p} \right) 
\cosh \left( \be {1+\g\over2} + \mu 
- {1\over2} \ln {{1-p}\over p}
\right)\right]\cr 
    & b=\ln 
{{\cosh \left(\be {1+\g\over2} -\mu
+ {1\over2} \ln {{1-p}\over p} \right)} 
\over
{\cosh \left( \be {1+\g\over2} +\mu
- {1\over2} \ln {{1-p}\over p}\right)}}
 + \be (1-\g)
}}\right.
}
It is possible to show that b is a monotonic decreasing function of $\mu$
with $-2\g \be \le b \le 2 \be$, so that we can directly use $b$ as 
variational parameter in order to realize the minimum in \free .

The effective hamiltonian $\tilde{H}_N$ is indeed associated 
to a pure $2d$ Ising model
with unitary strength couplings along the vertical bonds, and 
with alternated rows of unitary and $b\over{2\be}$ strength couplings.
This model can be solved by trivially
generalizing the Onsager solution 
and it is mapped into the problem of diagonalizing
a collection of $2 \times 2$ matrices.
In the thermodynamic limit $N \to \infty$
the total free energy \free\ reads 
$$
f(\be,\g)=-{\g+p(1-\g)\over2}-{1\over{2\be}}\ln\left[{4 p^p (1-p)^{1-p} } 
\sinh(\be(1+\g)) \sinh(2\be) \right] +
$$
$$
-\min_b\left[ {b\over{4\be}}(1-2p)-{1-p\over{2\be}}
\ln  (\e{b+2\be\g}-1) -{p\over{2\be}} \ln (\e{2\be-b}-1)  \right. +
$$
\eqn\freeb{
\left. +{1\over{4\pi\be}}\int_0^\pi dq \ln \lambda(q,b)  \ \right]
}
where 
$\lambda(q,b)$ indicates the maximum eigenvalue in modulus of the product
of the two matrices
$$
\left\{ \eqalign{
& {\bf T}_\be (q)=\exp\left[\be^* ({\bf \tau}_z \cos q+{\bf \tau}_x \sin q)
\right]
\exp(-2\be{\bf \tau}_z)
\cr
& {\bf \tilde T}_b (q)=\exp\left[\be^* ({\bf \tau}_z \cos q+{\bf \tau}_x 
\sin q) \right]
\exp(-b{\bf \tau}_z)
}\right.
$$
where $\be^*=-\ln\tanh\be$, and ${\bf \tau}_x$, ${\bf \tau}_z$ are Pauli 
matrices.
After some trivial algebra one gets:
\eqn\lam{
\lambda(q,b)  =t(q,b)+\sqrt{t(q,b)^2-1}
}
with:
$$
t(q,b)=2 \cos^2 q \ {\sinh b\over\sinh2\be} \
-\ 2 \cos q \ \cosh2\be \ {\sinh(2\be{+}b)\over\sinh^2 2\be} \ +
$$
\eqn\t{
+ \ {\cosh^2 2\be \ \cosh(2\be{+}b) \ + \ \cosh(2\be{-}b) \over\sinh^2 2\be}
}

The minimum of \freeb\ is realized for $b=b^*$ and it 
is achieved by looking for the zero of its derivative. One has the 
self-consistent equation for $b$:
\eqn\der{\left[
1+{2p\over{\e{2\be-b}-1}}-{2(1-p)\over{1-\e{-2\be\g-b}}}
+{1\over{4\pi}}\int_0^\pi dq {{\partial{t(q,b)} \over {\partial b}}
\over{\sqrt{t(q,b)^2-1}}}\right]_{b=b^*}=0
}
When $p\to 0$ the model has to reduce to the standard Ising
model and in fact, the previous formula leads to $b^* \to 2\be$, 
while in other limit case $p\to 1$ one has $b^* \to -2\g\be$.

Equation \der\ is an ordinary equation in $b$, nevertheless,
it cannot be explicitly solved so that we are not able
to give a compact expression for $b^*$ in terms of $T$, $\g$ and $p$.
However,
\der\ and then \freeb\ can be numerically computed with the necessary
precision in order to fully investigate the model.
Furthermore, at $ T=0$ while 
we don't have the complete solution of \der\
we are able to derive the leading terms of $b^*$ and 
to compute the ground state energy $U_0$.
We find out that $U_0$ has different linear behaviours in $p$
depending on $\g$ 
$$
U_0= -2 + ( 1 - {|1-\g| \over2} ) p 
$$

The $T=0$ entropy $S_0$ can only 
be computed numerically and
it is shown in \entropy . Unlike the energy $U_0$, $S_0$ 
equals a constant function of $p$ for any of the three choices of $\g$.
It is always zero for $\g<1$ and  $S_0\ge0$  for $\g=1$.
In the case $\g>1$ one can show that the entropy becomes positive
for $p>\tilde{p}$, where $\tilde{p}<{3\over4}$, and 
$S_0(p={3\over4})\simeq 0.01$.

\newsec{Transitions}

Let us stress again that formulae \freeb\ - \der\ represent the solution
of the model, and, in principle, all the informations about it 
can be derived from them. Fortunately, even if we are unable to give
an explicit expression of $b^*$ starting from \der , 
we can easily obtain some analytic results. For instance,
in this section we find the conditions that yield to a second order phase
transition.

Looking carefully at \freeb\ - \der , one can realize
that the mechanism of the usual Onsager transition is preserved:
the discontinuity occurs when $b^*$, the zero of \der ,
nullifies also the argument of the square root in \lam ,
i.e. when 
\eqn\tzero{
t(q,b^*)=1
} 
(the case $t(q,b^*)=-1$ is not possible 
since $t(q,b)\ge1$ for $\forall q$ and $\forall b$).
In other terms, one has to find out the solution $b^*$ of a system
of two equation, \der\ and \tzero . Obviously this solution can exist
only for certain values of $\be$, $\g$ and $p$.

A direct inspection of \t\ shows that \tzero\ can be satisfied
only for the specific choices $q=0$ or $q=\pi$, and it determines
the existence of two distinct transition lines in the $p-T$ phase diagram
at fixed $\g$, (see \phases\ ).
The first line ends on the $p=0$ (pure Ising model) axis at 
the Onsager critical temperature, so that in the following we will refer to 
this transition line as the ferromagnetic one. 
The second line exists only for $\g\ge1$ and it ends
on the $p=1$ axis in correspondence of the critical temperature 
separating the antiferromagnetic phase from the paramagnetic one;
for this reason we will call it the antiferromagnetic line.

After some trivial algebra \tzero\ reduces to
\eqn\trans{
\sinh(2\be+b^*)=\pm 2 {\cosh2\be\over\sinh^2 2\be}
}
where the sign $+$ correspond to the ferromagnetic line ($q=0$), 
and the  sign $-$ to the antiferromagnetic one ($q=\pi$).

In the limit case $p=0$ ($b^*=2\be$), the \trans\ recovers 
the well-known result $\sinh(2\be)=1$, while in the other limit case
$p=1$ ($b^*=-2\be\g$) the transition is present when
$\sinh (2\be(1-\g))=-2{\cosh 2\be \over \sinh^2 2\be}$, i.e.
at finite temperature when $\g>1$ and at zero temperature 
for the 'fully frustrated model' ($\g=1$).

From a practical point of view, in order to compute numerically 
the transition lines which are showen in \phases ,
it is convenient to solve \tzero\ with respect to $b^*$
\eqn\bstar{
b^*= \pm \ \rm{arc \, sinh} 
\left( 2{\cosh2\be\over\sinh^2 2\be}\right)- 2\be
}
Then, keeping $\g$ fixed and substituting $b^*$ into \der , one obtains 
the two transition lines $p(T)$, which can be easily
computed by standard numerical algorithms.

\newsec{Phases}
In the previous section we have seen that \trans\ gives
the known critical temperatures of the non random models 
($p=0$ or $p=1$). Another preliminary information about the behaviour
of the model comes from the observation that the
right hand side of \trans\ goes to zero in the limit $T\to0$, so that
the ferromagnetic and the antiferromagnetic lines must coincide
when they end on the $T=0$ axis.
Studying the leading terms of \der\ close to $T=0$, one finds that 
this coinciding point is at $p=1$ for $\g =1$ and
at $p={3\over4}$ for $\g>1$.

The full description of the different behaviours can be derived 
computing the transition lines in the $p - T$ phase diagram 
at varying $\g$, as seen in the previous section.
The following four scenarios listed in \phases , are obtained.

For $\g<1$ (see \phases a, where $\g=0.8$) 
only the ferromagnetic line is present,
separating two well-known phases: 
a ferromagnetic phase at low temperature, and a paramagnetic
phase at high temperature, exactly as for the Onsager non random model.
In fact the antiferromagnetic random couplings are too weak with respect to
the ferromagnetic ones, 
so that they are not able to change the structure of the phases 
of the pure model.

When $\g=1$ (\phases b) the scenario is quite similar 
to the previous one,
apart from the fact that the ferromagnetic line reaches the axis
$T=0$ at $p=1$, in correspondence of the $T=0$ transition
of the fully frustrated model.
Notice that the antiferromagnetic transition line is still absent.
In fact, the antiferromagnetic couplings have the same strength 
of the others but, for any $p<1$, their number is lower than the
number of  horizontal ferromagnetic couplings  so that
the ferromagnetic order prevails at low temperature. 

The most interesting situation corresponds to the choice $1<\g<2$
(\phases c, where $\g=1.2$). First of all notice that both the 
transition lines are present. The first line starts on the $p=0$ axis
and it ends at  $p={3\over4}$ on the $T=0$ axis and it delimitates the low
temperature ferromagnetic region.  The second line starts on the $T=0$ axis 
at $p={3\over4}$ ending on the $p=1$ axis, and it delimitates the
antiferromagnetic phase.
Outside this two regions there is a non-magnetic phase, but notice that
this one has a narrow tongue dividing the magnetic regions
and reaching the $T=0$ axis at $p={3\over4}$. 
As a consequence, if one fixes the probability between
${3\over4} < p < {3\over4}+\delta p(\g)$, where $\delta p(\g)$ is a small
but finite number depending on $\g$,
one can observe three different second order phase transitions
varying the temperature $T$.
The transitions separate four phases; 
starting from low temperatures, the first is ferromagnetic,
the third is antiferromagnetic, the fourth is an ordinary 
paramagnetic phase while the second is a low temperature paramagnetic phase.
In \cal\ it is shown the specific heat $C$
as a function of $T$ at fixed $\g=1.2$ and $p=0.82$, 
computed starting from the numerical solution of \freeb\ and \der .
In this case the specific heat $C$ exhibits three distinct peaks 
next to $T\simeq0.299$, $T\simeq0.349$ and $T\simeq0.545$.
Indeed we need to magnify the picture since it is necessary to compute
\freeb\ with a great precision in order to show a certain growth of $C$
around its discontinuity.

The appearance of a low temperature paramagnetic phase between the antiferro 
and the ferromagnetic ones represents an interesting
peculiarity of this model. In particular, it happens
at relatively low temperature and with an extremely narrow width.
These are the main features that persuade us to guess a glassy nature
for this paramagnetic phase.
Moreover in our constrained annealed model, 
as seen at the end of section 3,
the region at low temperature with an unphysical solution 
(negative zero temperature entropy $S_0$) do not reach
the critical transition point ($p={3\over4}$ , $T=0$) 
where the 'glassy' paramagnetic phase ends.

The description of the different scenarios is completed
with the case $\g>2$ (\phases d, where $\g=3$). 
The structure of the phase diagram is similar
to the previous one with the difference that the 
narrow tongue between the ferro and the
antiferro phases is suppressed.
As a consequence for $p<{3\over4}$ 
we only have the ferro and the para phases while for 
$p>{3\over4}$ we only have the antiferro and the para 
phases. When $\g\to\infty$ the temperature of  
end point on the $p=1$ axis goes to infinite.

\newsec{Conclusions}

The surprising feature of our model is that for some choices of 
the parameters $\g$ and $p$ the magnetic phases
are separated by a low temperature paramagnetic phase.
We do not expect any long distance magnetic order in this phase i.e.
$\avd{ {<}\sigma_{i,j} \sigma_{i,j+k} {>} } 
=\avd{ {<}\sigma_{i,j} \sigma_{i+k,j} {>} }=0$ in the limit $k\to \infty$
but we expect that  
$\avd{ {<} \sigma_{i,j} \sigma_{i,j+k} {>}^2 } >0$ 
and $\avd{ {<} \sigma_{i,j} \sigma_{i+k,j} {>}^2 } >0$ in the same limit.
Our proposal is that these last two quantities properly characterize the
glassy-like paramagnetic phase and they should vanish in 
the high temperature paramagnetic phase.
Unfortunately, it is well known that in two
dimensional models the computation of
long distance correlation is a difficult task and, in our case,  
the computation also involves averages over the disorder
making the situation even more complicated.
We think that some work can be made in this direction but it will demand 
much technical effort so that our claim that the low
temperature paramagnetic phase is somehow a glassy phase is, at this point, 
more a conjecture than an established fact.

Two more questions remain to be answered. 
The first is the most relevant: what is the role of the annealed approximation
in the qualitative features of the phase diagram? Or better,
is the new phase a mere consequence of the annealed approximation?
In this case, our fixed frustration model would have an interest in itself 
but it would not be a good approximation of the quenched one.
We think that this question can only be answered by direct 
Montecarlo simulation.

The second question is: in this model
the frustrated plaquettes appears only in couples, what is the role 
of this special correlation? 
To be more specific, a model in which all vertical bonds equals the unity
while the horizontal are independent random variables 
which take two possible values of opposite sign
would have the same qualitative behaviour?
We cannot approximate such a model by our fixed frustration technique so that
also this question can only be answered by Montecarlo simulation, 
nevertheless, 
we are convinced that the answer should be positive.
In fact, the special nature of correlation between plaquettes is short 
ranged and there is not reason why it should affect 
the long range behavior of the system.

In conclusion we would like to stress that in spite of the very partial 
results 
contained in this paper and of the many unsolved questions 
this work sheds some light on the very important point
of the existence of a finite temperature glassy phase for $d=2$ 
frustrated systems.
In fact, in the most restrictive interpretation of our result
we can still affirm that the fixed frustration annealed $d=2$ model
has low temperature glassy-like phase,
while in the most generous one we can say that a true glassy phase exists for
quenched random $d=2$ systems.

\bigskip
\bigskip
\bigskip
\noindent
{\bf Acknowledgements}
\bigskip

We acknowledge the financial support of the I.N.F.N., 
National Laboratories of Gran Sasso ({\it Iniziativa Specifica} FI11).
We thank Roberto Baviera for useful discussions.

\vfill\eject

\listrefs

\listfigs

\bye